\documentclass[12pt,titlepage]{article}
\usepackage{epsfig}
\usepackage{graphicx}
\usepackage{amsmath}

\def\beq{\begin{equation}}
\def\eeq{\end{equation}}
\def\bea{\begin{eqnarray}}
\def\eea{\end{eqnarray}}
\def\dCSA{N}

\def\J{{\widetilde a}}
\def\Inertia{{\cal I}}
\def\K{{\cal K}}
\def\E{{\cal E}}
\def\U{{\cal W}}

\def\Vsphere{{\varpi}}
\def\cosmo{{\widetilde\lambda}}

\begin{document}
\title{Thermodynamic stability of asymptotically anti-de Sitter rotating black holes in higher dimensions}

\author{{Brian P. Dolan}\\
{\em Department of Mathematics, Heriot-Watt University}\\ 
{\em Edinburgh, EH14 4AS, U.K.}\\
[5mm]
{\em Maxwell Institute for Mathematical Sciences, Edinburgh, U.K.}\\
[5mm]
{\em Email:} {\tt B.P.Dolan@hw.ac.uk}
}

\maketitle
\begin{abstract}
Conditions for thermodynamic stability of asymptotically anti-de Sitter
rotating black holes in $D$-dimensions are determined. 
Local thermodynamic stability requires not only positivity conditions on the specific heat and the moment of inertia tensor but it is also necessary that the adiabatic compressibility be positive. 
It is shown that, in the absence of a cosmological constant, neither rotation nor charge is sufficient to ensure full local thermodynamic stability of a black hole.

Thermodynamic stability properties of anti-de Sitter Myers-Perry black holes are investigated for both singly spinning and multi-spinning black holes. 
Simple expressions are obtained for the specific heat and moment of inertia
tensor in any dimension.  An analytic expression is obtained for the boundary of the region of parameter space in which such space-times are thermodynamically stable.

\bigskip

\

\noindent
\rightline{PACS nos: 04.50.Gh; 04.60.-m;  04.70.Dy}
\end{abstract}

\section{Introduction}

Ever since Hawking's discovery \cite{Hawking} that black holes have a temperature associated with them, and will radiate thermal energy when isolated, the fascinating topic of black hole thermodynamics has been a major focus
of research.  Schwarzschild black holes have a negative specific heat and are
therefore thermodynamically unstable but, for anything other than very small mass black holes, the life-time is so large that the black hole persists
for time scales greater than or of the order of the age of the Universe 
and thermodynamic concepts can still be applied.
It is possible to make a black hole thermodynamically stable by introducing a 
negative cosmological constant $\Lambda$ and considering asymptotically anti-de Sitter (AdS) black holes, provided the magnitude of $\Lambda$ is large enough,
\cite{HawPag}.\footnote{They can also be stabilised by putting them in a finite volume cavity, \cite{York}.} 

The specific heat for an asymptotically flat black holes
can also be rendered positive by rotating
the black hole or by giving it an electric charge, 
but this does not in itself ensure full thermodynamic stability
as the moment of inertia can, and indeed does, become negative.  
This phenomenon persists
to space-times of higher dimension \cite{DFMRS}. 
In dimension $D>4$ there is more than one angular momentum and the 
moment of inertia is a tensor but this tensor
develops a negative eigenvalue in any region of parameter space for which
the corresponding specific heat is positive \cite{DFMRS}, \cite{BPDThermo}. 

The situation for asymptotically AdS black holes in $D$ dimensions is somewhat
different and, with the current interest in the AdS/CFT correspondence 
\cite{Maldacena}, such space-times are of considerable importance.
The literature here is not as comprehensive as for the asymptotically
de Sitter case,
though there has been significant progress
in specific cases \cite{MPS-AdS}-\cite{KRT}.  
In this paper a full description is given
of the local thermodynamic stability properties of asymptotically AdS black
holes in any dimension.
It is shown in \S\ref{sec:GeneralStability} that all isolated black holes 
suffer from a local thermodynamic instability unless a negative cosmological
constant is introduced.
In \S\ref{sec:MP} a well known class of rotating asymptotically AdS black holes, 
asymptotically AdS Myers-Perry black holes, is studied in detail and the thermodynamically stable region of parameter space is mapped out for all space-time dimensions.
Finally \S\ref{sec:conclusions} summarises our results.

\section{General criteria for stability \label{sec:GeneralStability}}

When a cosmological constant is present it was argued in \cite{KRT}
that the ADM mass of a black hole should be viewed, in a thermodynamic context, as the enthalpy $H$ of the thermodynamic system, rather than the heretofore
more common interpretation of internal energy $U$, and this will be the philosophy adopted here. The natural thermodynamic control parameters for the enthalpy are entropy $S$, angular momentum $J$, pressure $P$ and electric charge $Q$, so we write
\beq M=H(J,S,P,Q).\eeq
The thermal energy $U(J,S,V,Q)$, which depends not only on $S$, $J$, $Q$ but also on the volume $V$, is the Legendre transform of the enthalpy
\beq
U=H-PV.
\eeq
There are subtleties associated with the definition of the volume 
of a black hole, 
\cite{Hayward:1997jp} - \cite{Ballik:2013uia},
but thermodynamically volume can be defined
as the variable thermodynamically conjugate to the pressure \cite{KRT,BPDVolume}, 
\beq
V=\left.\frac{\partial M}{\partial P}\right|_{J,S,Q},
\eeq
where the pressure is produced by the cosmological constant
\beq P=-\frac{\Lambda}{8\pi}.\eeq
The introduction of a thermodynamic pressure and volume in black hole physics 
has generated a lot of activity on the black hole equation of state
recently, for a review of the current situation see \cite{AKMS}. 

Local thermodynamic stability requires that the internal energy $U$ be
a convex function \cite{Callen}. 
Equivalently the Legendre transform of $U$, 
\beq \E(\Omega,T,P,\Phi):=U -\Omega_i J^i - TS + PV -\Phi Q
=M -\Omega_i J^i - TS -\Phi Q,\eeq
should be a concave function.\footnote{In $D$ space-time dimensions there is more than one
angular momentum, labelled by the index $i$ which runs from 1 up to $\dCSA$,
where $\dCSA$ is the dimension of the Cartan sub-algebra of $SO(D-1)$.}  
We shall refer to $(J^i,S,V,Q)$ as extensive variables and $(\Omega_i,T,P,\Phi)$ as intensive variables. This classification is motivated by the canonical dimensions of these variables, as the dimensions of $(J,S,V,Q)$ depend on $D$ while those of
$(\Omega_i,T,P,\Phi)$ do not, as shown in the table below (with Newton's constant and the speed of light set to unity),

\medskip

\setlength{\tabcolsep}{15pt}
\centerline{\begin{tabular}{| l| c| c| }
\hline
Thermodynamic Variable & Dimension \\
\hline
 Mass, $M$   &  $D-3$ \\
  Entropy, $S$ (area)  & $D-2$ \\
  Angular momenta, $J^i$ & $D-2$ \\
 Volume, $V$ & $D-1$ \\
 Electric Charge, $Q$ & $D-3$\\ 
 Temperature,  $T$ & $-1$ \\
 Angular velocity,  $\Omega_i$ & $-1$ \\
 Pressure, $P$ ($\Lambda$) & $-2$ \\
 Electric potential, $\Phi$ &  $0$ \\
\hline
\end{tabular}}

\bigskip

Let 
\beq x_A=\frac{\partial U}{\partial X^A}=(\Omega_i,T,-P,\Phi), \eeq
with $A=1,\ldots,\dCSA +3$, denote the intensive variables and
\beq X^A=-\frac{\partial \E}{\partial x_A}=(J^i,S,V,Q)\eeq 
denote the extensive variables.

In the grand canonical ensemble the volume is usually kept fixed 
and the other thermodynamic control parameters are the intensive ones. 
In ordinary thermodynamics scaling arguments would imply that
Legendre transforming to make {\it all} control parameters intensive results in a
vanishing thermodynamic potential, $\E=0$, \cite{Callen} (this is essentially the content of the 
Gibbs-Duhem relation). For black holes however this is not 
the case and $\E(x_A)$ is not trivial, in fact
$\E$ is related to the Euclidean action $I_E$ by \cite{GPPI}
\beq \E=T I_E.\eeq 
We shall call the ensemble with thermodynamic potential $\E$ 
the extended canonical ensemble and $\E$ the extended free energy.

For asymptotically flat black holes $P=0$ and the grand canonical and extended
canonical ensembles coincide, but for asymptotically AdS black holes they do not.
(The relation between the microcanonical, the canonical and the grand canonical ensembles for asymptotically flat black holes 
was elucidated in \cite{BPDThermo}.)

A full analysis of local thermodynamic stability
requires determining the region of parameter space in which all
of the the eigenvalues
of the Hessians, either
\beq \U_{AB}=\frac{\partial^2 U}{\partial X^A \partial X^B}  
\label{eq:WAB}\eeq
or 
\beq \K^{AB}=-\frac{\partial^2 \E}{\partial x_A \partial x_B},  \eeq
are positive.  It does not matter which one is used, since it is a 
standard fact of Legendre transforms that $\K=\U^{-1}$, we shall start by
focusing on $\U_{AB}$.  

It was shown in \cite{DFMRS} that all asymptotically flat, electrically neutral 
black holes are unstable. We present here a different derivation of this result which has the advantage of generalising it 
to include electric charge.  Using the above dimensions in the Smarr relation \cite{KRT, Smarr} gives 
\beq
(D-3)M= (D-2)\,\mathbf{\Omega. J}+(D-2)TS  -2PV+(D-3) Q\Phi.
\label{eq:Smarr}
\eeq
 With $M=U+PV$ this can be re-arranged, for $D\ge 4$, as
\bea
U&=& \frac{(D-2)}{(D-3)}\,\mathbf{\Omega. J} +\frac{ (D-2)}{(D-3)}TS
-\frac{(D-1)}{(D-3)}PV+ Q\Phi\nonumber\\
&=&\sum_B c_B  X^B x_B,
\eea
with $c_A$ the appropriate $D$-dependent constants.
Differentiating this
\bea
x_A=\frac{\partial U}{\partial X^A} &=& c_A x_A + \sum_B c_B \U_{AB} X^B\nonumber \\
\Rightarrow  \qquad \sum_B \U_{AB} \bigl(c_B  X^B\bigr)  &=& (1-c_A)x_A,
\label{eq:UcX}
\eea
(there is no sum over $A$ here, for the rest of this section we suppress the summation convention and
explicitly show all summations).
The vector on $X$-space with components ${\cal D}^A=c_A X^A$ essentially 
represents the response of the system to a re-scaling.
Using (\ref{eq:UcX}) we can construct the quadratic form
\bea
\sum_{A,B}\U_{AB}{\cal D}^A {\cal D}^B &=&
\sum_A c_A (1-c_A) x_A X^A \hfill\nonumber\\
&=&\frac{(D-2)}{(D-3)^2}\left(-\mathbf{\Omega. J} -TS +2\frac{(D-1)}{(D-2)} P V
 \right).\label{eq:QuadraticForm}
\eea
Clearly $\sum_{AB}\U_{AB}{\cal D}^A {\cal D}^B<0$ if $P=0$, heralding
a thermodynamic instability for any asymptotically flat
black hole in $D>3$ space-time dimensions, regardless of charge or rotation.  However, provided $V>0$,  
a positive $P$ (negative $\Lambda$)
can remove this particular instability if the $PV$ term is of sufficient
magnitude to outweigh the $TS$ and $\mathbf{\Omega. J}$ terms
on the right-hand side of (\ref{eq:QuadraticForm}).\footnote{The quadratic form $\U_{AB}$ can be interpreted as a metric on configuration space,
the Ruppeiner metric, \cite{Ruppeiner}, 
and stability requires that this metric be positive definite, or at least non-negative.
In ordinary thermodynamics, the Gibbs-Duhem relation implies $c_A=1$, for all $A$, 
and ${\cal D}^A$ is always a null vector, indicating a direction of neutral stability.} 
This is a very powerful result, it shows that the only way of making an isolated black hole locally thermodynamical stable, without imposing constraints by fixing variables, is to introduce a positive pressure. Rotation and/or electric charge alone cannot do the job unless they are fixed and not allowed to vary,\footnote{Stabilising the black hole by fixing $J$ and/or $Q$ can lead to very interesting phenomena,
for example there is a phase transition between large and small
black holes when $Q$ is fixed \cite{CEJM} or when $J$ is fixed \cite{CCK}.
The equation of state leads to van der Waals type critical points,
with mean field exponents \cite{KM,PdV} and there can even be a triple point
\cite{AKMS3} and re-entrant phase transitions \cite{AKM}.} 
but with no constraints there cannot be local thermodynamical 
stability unless $P>0$.
The case $P=Q=0$ was explored in \cite{BPDThermo}, 
and the $P>0$ scenario will be analysed in some detail in this work, 
but still with $Q=0$.

It is shown in appendix \ref{app:Stability} that, for $Q=0$,  the mathematical
requirement of positivity of $\U$ is equivalent to following three 
perfectly reasonable physical statements (assuming that 
the temperature and the thermodynamic volume are positive):
\begin{itemize}
\item the specific heat at constant $\Omega_i$ is positive,
\beq
C_\Omega = T\left.\frac{\partial S}{\partial T}\right|_{\Omega,P}>0;
\eeq
\item the isentropic moment of inertia tensor
\beq
\Inertia^{ij}
=\left.\frac{\partial J^i}{\partial \Omega_j}\right|_{S,P}
=\left.\frac{\partial J^j}{\partial \Omega_i}\right|_{S,P}
\eeq
is a positive matrix;
\item the adiabatic compressibility is positive,
\beq
\kappa=-\left.\frac{1}{V}\frac{\partial V}{\partial P}\right|_{J^i,S}>0.
\eeq
\end{itemize}
All perfectly reasonable physical requirements.

In the next section we shall examine these conditions in the specific
case of asymptotically AdS Myers-Perry black holes in $D$-dimensions
and determine the region of parameter space for which these space-times
are thermodynamically stable.

\section{Asymptotically anti-de Sitter Myers-Perry black holes.\label{sec:MP}}

\subsection{The metric and thermodynamic variables \label{sec:MPmetric}}

Rotating black holes in $D$-dimensions must be treated slightly
differently for even and odd $D$
because the rotation group  $SO(D-1)$, 
acting on the event horizon which is assumed to have
the topology of a $(D-2)$-dimensional sphere, has different characterisations 
of angular momenta in the even and odd 
dimensional cases.
The Cartan sub-algebra has dimension $\frac{D-2}2$ for even $D$ and $\frac{D-1}2$ for odd $D$ so a general state of rotation is specified by 
$\frac{D-2}2$ independent angular momenta in even $D$ and $\frac{D-1}2$
in odd $D$.  Let $\dCSA=\left\lfloor \frac{D-1}2 \right\rfloor$, the integral part
of $\frac{D-1}2$, be the dimension of the Cartan sub-algebra of $SO(D-1)$,
then there are $\dCSA$ independent angular momenta $J^i$,
$i=1,\ldots,\dCSA$.  It is notationally convenient to define $\epsilon=\frac{1+(-1)^D}{2}$, so $\epsilon=1$
for even $D$ and $\epsilon=0$ for odd $D$, and then \beq \dCSA=\frac{D-1-\epsilon}2.\eeq

In this notation the unit $(D-2)$-dimensional sphere can be described in
terms of Cartesian co-ordinates ${\tt x}_\alpha$ in
${\bf R}^{D-1}$ by
\beq \sum_{\alpha=1}^{D-1} {\tt x}_\alpha^2 =1,\eeq
and we write this as
\beq \sum_{i=1}^\dCSA \rho_i^2 + \epsilon y^2=1,\eeq
where ${\tt x}_{2i-1}+i{\tt x}_{2i} = \rho_i e^{i\phi_i}$, $i=1,\ldots,\dCSA$, are complex co-ordinates
for both the even and odd cases while $y={\tt x}_{D-1}$ is only necessary for even $D$.
$\rho_i$, $\phi_i$ and $y$ are then (over complete) co-ordinates that can be used
to parameterise the $(D-2)$-sphere and, for the black hole, $J^i$ are angular
momenta in the $({\tt x}_{2i-1},{\tt x}_{2i})$-plane.

The first rotating black solutions to Einstein's equations
in dimension greater than four were
the asymptotically flat solutions of Myers and Perry \cite{MP}.
Rotating black holes in $5$-dimensions with a cosmological constant,
$\Lambda$, were constructed in \cite{HHT-R} and the generalisation to
the $D$-dimensional
metric was found in \cite{GPP}:
these are solutions of Einstein's equations with Ricci tensor
\beq
R_{\mu\nu}=\frac{2\Lambda}{(D-2)}g_{\mu\nu}.
\eeq 
We shall focus on $\Lambda\le 0$ here, as the thermodynamics is then
better understood, and for notational convenience we define
\beq \lambda=-\frac{2\Lambda}{(D-1)(D-2)}\ge 0.\eeq
The line element in \cite{GPP} can then be expressed, in
Boyer-Linquist co-ordinates, as\footnote{The form given here differs
slightly from that in \cite{GPP} in that our ordinates, $t$ and $\phi_i$,
are related to those of \cite{GPP}, $\tau$ and $\varphi_i$, by 
$d\tau=dt$ and $d\phi_i = d\varphi_i -\lambda a_i dt$.}
\bea
d s^2&=&-W(1+\lambda r^2)dt^2 +\frac {2\mu }U
\left(W dt-\sum_{i=1}^\dCSA  \frac{a_i \rho_i^2d\phi_i}{\Xi_i}\right)^2
\nonumber\\
&& +\left(\frac {U }{Z-2\mu}\right)d r^2 + \epsilon\, r^2 d y^2 
 +\sum_{i=1}^\dCSA\left(\frac{r^2+a_i^2}{\Xi_i}\right)(d\rho_i^2+\rho_i^2d\phi_i^2 )\\
&&-\frac{\lambda}{W(1+\lambda r^2)}
\left(\sum_{i=1}^\dCSA \left(\frac{r^2+a_i^2}{\Xi_i}\right)\rho_i d\rho_i +
\epsilon r^2 ydy\right)^2,
\label{eq:GPPds}
\nonumber\eea
where 
\beq \Xi_i = 1-\lambda a_i^2\eeq
and the functions $W$, $Z$ and $U$ are 
\bea W&=& \epsilon y^2+\sum_{i=1}^\dCSA \frac{\rho_i^2}{\Xi_i}\nonumber\\
Z&=&\frac{(1+\lambda r^2)}{r^{2-\epsilon}}\prod_{i=1}^\dCSA(r^2+a_i^2)\\
U&=&\frac{Z}{1+\lambda r^2}\left(1-\sum_{i=1}^\dCSA\frac{a_i^2\rho_i^2}{r^2+a_i^2} \right).\nonumber\eea
The $a_i$ are rotation parameters in the $({\tt x}_{2i-1},{\tt x}_{2i})$-plane,
restricted to 
\beq a_i^2<1/\lambda,\label{eq:Xi-restriction}\eeq 
and $\mu$
is a mass parameter.

Many of the properties of the space-time with line element
(\ref{eq:GPPds}) were described in \cite{GPP}. 
There is an event horizon at $r_h$, the largest root of $Z-2\mu=0$, 
so
\beq 
\mu=\frac{(1+\lambda r_h^2)}{2r_h^{2-\epsilon}}\prod_{i=1}^\dCSA(r_h^2+a_i^2),
\label{eq:mudef}\eeq
with area
\beq 
{\cal A}_h=\frac{\Vsphere}{r_h^{1-\epsilon}}\prod_{i=1}^\dCSA\frac{r_h^2+a_i^2}{\Xi_i}\,,
\label{eq:Adef}
\eeq
where $\Vsphere$ is 
is the volume of the round unit $(D-2)$-sphere,
\beq
\Vsphere= \frac{2\pi^{\frac{(D-1)}{2}}}{\Gamma\left(\frac{D-1}{2} \right)}\,.
\eeq

The Bekenstein-Hawking entropy is
\beq 
S=\frac{\Vsphere}{4 r_h^{1-\epsilon}}\prod_{i=1}^\dCSA\frac{r_h^2+a_i^2}{\Xi_i}
\label{eq:Sdef}\eeq
and the Hawking temperature is, with $\hbar=1$,
\beq
T=\frac{r_h}{2\pi}(1+\lambda r_h^2)\sum_{i=1}^\dCSA\frac{1}{r_h^2 + a_i^2} + \frac{(2-\epsilon)(\epsilon\lambda r_h^2-1)}{4\pi r_h}.
\label{eq:THawk}
\eeq

The  angular momenta and the ADM mass, $M$, of the black hole are related to the metric parameters via
\bea  J_i &=& 
\frac{\mu\,\Vsphere a_i}{4\pi\Xi_i\prod_j \Xi_j},   \label{eq:Jdef} \\
M&=&\frac{\mu\,\Vsphere}{8\pi\prod_j \Xi_j} 
\left(D-2+2\lambda\sum_{i=1}^\dCSA\frac{a_i^2}{\Xi_i}\right)
\label{eq:Mdef}\\
&=&\frac{S}{4\pi r_h}(D-2)(1+\lambda r_h^2)+\lambda\sum_{i=1}^\dCSA J_i a_i ,\nonumber
\eea 
while the angular velocities are
\beq \Omega_i=\frac{(1+\lambda r_h^2)a_i}{(r_h^2+a_i^2)}.\eeq

\subsection{Thermodynamic potentials \label{sec:EnthalpyFreeEenrgy}}

Following \cite{KRT} the ADM mass of the black hole (\ref{eq:Mdef}) is identified with the enthalpy 
\beq
M=H(J,S,P).
\eeq
The Legendre transform to purely intensive variables,
the extended free energy
\beq
\E(\Omega,T,P)=H -TS-{\mathbf \Omega}.{\mathbf J}=U+PV -TS-{\mathbf \Omega}.{\mathbf J},
\eeq
then has a simple form for asymptotically AdS Myers-Perry black holes.
\beq
\E=\frac{S}{4\pi r_h} (1-\lambda r_h^2).
\label{eq:GrandCan}
\eeq 
The observation that this is negative for $\lambda r_h^2>1$, and
hence less than the corresponding potential for pure AdS space-time with $S=0$
and no black hole, is the origin of the Hawking-Page phase transition \cite{HawPag}.

Using the Smarr relation (\ref{eq:Smarr}) one finds that,
for any electrically neutral black hole,
 the internal energy, $U=M-PV$, is related to the Euclidean action
$\E=M-TS-{\mathbf \Omega. J}$
\beq U =\frac{(D-2)\E +M}{2},\label{eq:UEM}\eeq
and this is easily checked explicitly for AdS Myers-Perry solutions.
 
\subsection{Thermodynamic volume \label{sec:Volume}}

The thermodynamic volume, $V$, is defined as the variable 
thermodynamically conjugate to $P$,
\beq \label{eq:dMVolume}
V=\left.\frac{\partial M}{\partial P}\right|_{S,J}
= \frac{16\pi}{(D-1)(D-2)}\left.\frac{\partial M}{\partial \lambda}\right|_{S,J}.
\eeq
This was evaluated for electrically neutral asymptotically AdS Myers-Perry black holes
in \cite{CGKP}, using the
Smarr relation (\ref{eq:Smarr}) with $Q=0$, and the expressions 
given in \S\ref{sec:MPmetric}
for $M$, $T$,
$S$, $\Omega_i$ and $J^i$.  
Alternatively one can use (\ref{eq:dMVolume}) directly,
 \cite{BPDThermo}. 
In either case the result is
\bea \label{eq:Volume}
V&=&\frac{r_h {\cal A}_h}{D-1}\left\{1+\frac{(1+\lambda r_h^2)}{(D-2) r_h^2}\sum_{i=1}^\dCSA \frac{a_i^2}{\Xi_i} \right\}\\
&=&\frac{r_h {\cal A}_h} {D-1}+\frac{8\pi}{(D-2)(D-1)} \sum_{i=1}^\dCSA a_i J_i\nonumber.
\eea
Clearly $V>0$, since $a_i$ and $J^i$ are always of the same sign.

With the substitution $\lambda\rightarrow - g^2$ (\ref{eq:Volume}) 
also agrees with the black hole thermodynamic volume quoted in \cite{MPdS} 
for $\Lambda > 0$, again determined by assuming the Smarr
relation holds.

\subsection{Temperature \label{sec:Temperature}}

To discuss the various other thermodynamic quantities it will prove useful 
to work with 
dimensionless variables 
\beq \J_i=\frac{a_i}{r_h} \qquad \hbox{and} \qquad \cosmo=\lambda r_h^2,
\label{eq:atildecosmodef}
\eeq
in terms of which
the Hawking temperature (\ref{eq:THawk})
can be re-expressed as
\beq
T=\frac{D-3-2\Sigma_1 +\cosmo(D-1-2\Sigma_1)}{4\pi r_h},
\eeq
where 
$\Sigma_1 :=\sum_{i=1}^\dCSA \frac{\J_i^2}{1+\J_i^2}$.
There is of course a restriction on $\J_i$ and $\cosmo$ that defines the region of parameter space 
where $T>0$.  In general $T$ will vanish on a $N$-dimensional hypersurface 
in $(\J_i,\cosmo)$-space given by the zero locus of the polynomial
\beq
 (D-3)\prod_{i=1}^\dCSA (1+\J_i^2) -2\sum_{i=1}^\dCSA \J_i^2 \prod_{j\ne i}(1+\J_j^2) +\cosmo \bigl\{(D-1)\prod_{i=1}^\dCSA (1+\J_i^2) -2\sum_{i=1}^\dCSA \J_i^2 \prod_{j\ne i}(1+\J_j^2)\bigr\}.
\eeq

Rather than performing the most general 
analysis, which would get increasingly involved for larger and larger $D$, 
the discussion
here will be limited to special configurations in which $n$ of the $\J_i$ are non-zero and equal and the remaining $\J_i$ vanish,
this will prove sufficient to gain a good understanding of the
thermodynamics. So we choose 
\beq \J_1=\cdots = \J_n=\J,\qquad \J_{n+1}=\cdots=\J_\dCSA=0.
\label{eq:n-equal-a}\eeq
The case $n=1$ is called singly-spinning and the cases $n>1$ will be called
multi-spinning.

In these configurations, the $T=0$ hypersurface is given by
\beq
D-3 + (D-3-2n)\J^2 +\bigl(D-1 +(D-1-2n)\J^2 \bigr)\cosmo=0
\eeq
on which
\beq
\J^2=\frac{D-3 +(D-1)\cosmo}{2n-(D-3) +\bigl(2n-(D-1)\bigr)\cosmo}\,.
\eeq

We thus see that demanding positivity of $T$ puts the following restrictions on
$\J$:
\begin{enumerate}
\item If $n \le \frac{D-3}{2}$, no restriction on $\J^2$.
\item Even $D$, if $n=N=\frac{D-2}{2}$ and $\cosmo<1$, then $\J^2<\frac{D-3+(D-1)\cosmo}{1-\cosmo}$.
\item Odd $D$, if $n=N=\frac{D-1}{2}$, $\J^2<\frac{D-3 +(D-1)\cosmo}{2}$.
\end{enumerate}

Thus positivity of $T$ is only an issue for the configurations (\ref{eq:n-equal-a}) when the black hole is spinning in all possible planes and $n=N$.

\subsection{Specific Heat}

$C_\Omega$ can be evaluated straightforwardly and the details are left to appendix \ref{app:COmega}.  Here we just quote the result, 
\begin{equation}
  C_{\Omega} = - \frac{4 \pi r_h T S \left(D-2+2 \overline{\Sigma}_1 \right)}
{\left( D - 3 + 2 \overline{\Sigma}_1 - \cosmo (D - 1 + 2
\overline{\Sigma}_1) \right)},\label{eq:C_Omega}
\end{equation}
where 
$\overline{\Sigma}_1 :=\sum_{i=1}^p \frac{\J_i^2}{1-\J_i^2}$. 
In asymptotically flat Myers-Perry space times this reduces to the 
result 
\[ C_{\Omega} = - \frac{16 \pi^2 r_h^2 M T (D - 2 + 2 \overline{\Sigma}_1)}{(D - 2)
   (D - 3 + 2 \overline{\Sigma}_1)}\,, \]
in \cite{BPDThermo}. 

Local thermodynamic stability requires that $C_\Omega$
be positive and this only holds in a restricted region of the $(\dCSA+1)$-dimensional parameter space $(\J_i,\cosmo)$.  Assuming that the metric parameter $\mu$ is chosen so that there is a black hole and $r_h>0$, 
and that the condition (\ref{eq:Xi-restriction}) is satisfied so that $S>0$, then,
from (\ref{eq:C_Omega}),
\beq
\frac{C_\Omega}{S} \ge 0\quad\Leftrightarrow \quad
\frac{\bigl( D - 3 - 2 \Sigma_1 + \cosmo (D - 1 - 2\Sigma_1) \bigr)\bigl(D-2+2 \overline{\Sigma}_1 \bigr)}
{\bigl( D - 3 + 2 \overline{\Sigma}_1 - \cosmo (D - 1 + 2\overline{\Sigma}_1) 
\bigr)} \le 0.\label{eq:C_Omega-positivity}
\eeq
The boundary between regions of opposite sign is the locus of points where either the
numerator is zero or the denominator vanishes and 
$C_\Omega$ has a pole. Introducing the notation 
\bea
{\cal C}_{D,s}&=&(D-s)\prod_i^\dCSA(1+\J_i^2) - 2\sum_{k=1}^\dCSA \left(\J_k^2 
\prod_{i\ne k}^\dCSA(1+\J_i^2)\right)\nonumber \\
\overline{\cal C}_{D,s}&=&(D-s)\prod_i^\dCSA(1-\J_i^2) + 2\sum_{k=1}^\dCSA \left(\J_k^2 
\prod_{i\ne k}^\dCSA(1-\J_i^2)\right),\label{eq:C_s}
\eea
the zeros of $C_\Omega$ lie on hypersurfaces characterised by 
\beq \overline{\cal C}_{D,2}=0,
\eeq
or
\beq
{\cal C}_{D,3} +\cosmo\,{\cal C}_{D,1}=0
\eeq
(the latter is the $T=0$ hypersurface).
There are poles in $C_\Omega$ on the hypersurface characterised by
\beq \overline{\cal C}_{D,3} -\cosmo\,\overline{\cal C}_{D,1}=0. 
\label{eq:singular_surface}\eeq
It is curious that the zeros of $T$ and the poles of $C_\Omega$
are characterised by the same polynomial but
with $\J^2 \rightarrow -\J^2$ and $\cosmo \rightarrow -\cosmo$.

\begin{figure}[!ht]
\centerline{\includegraphics[width=8cm]{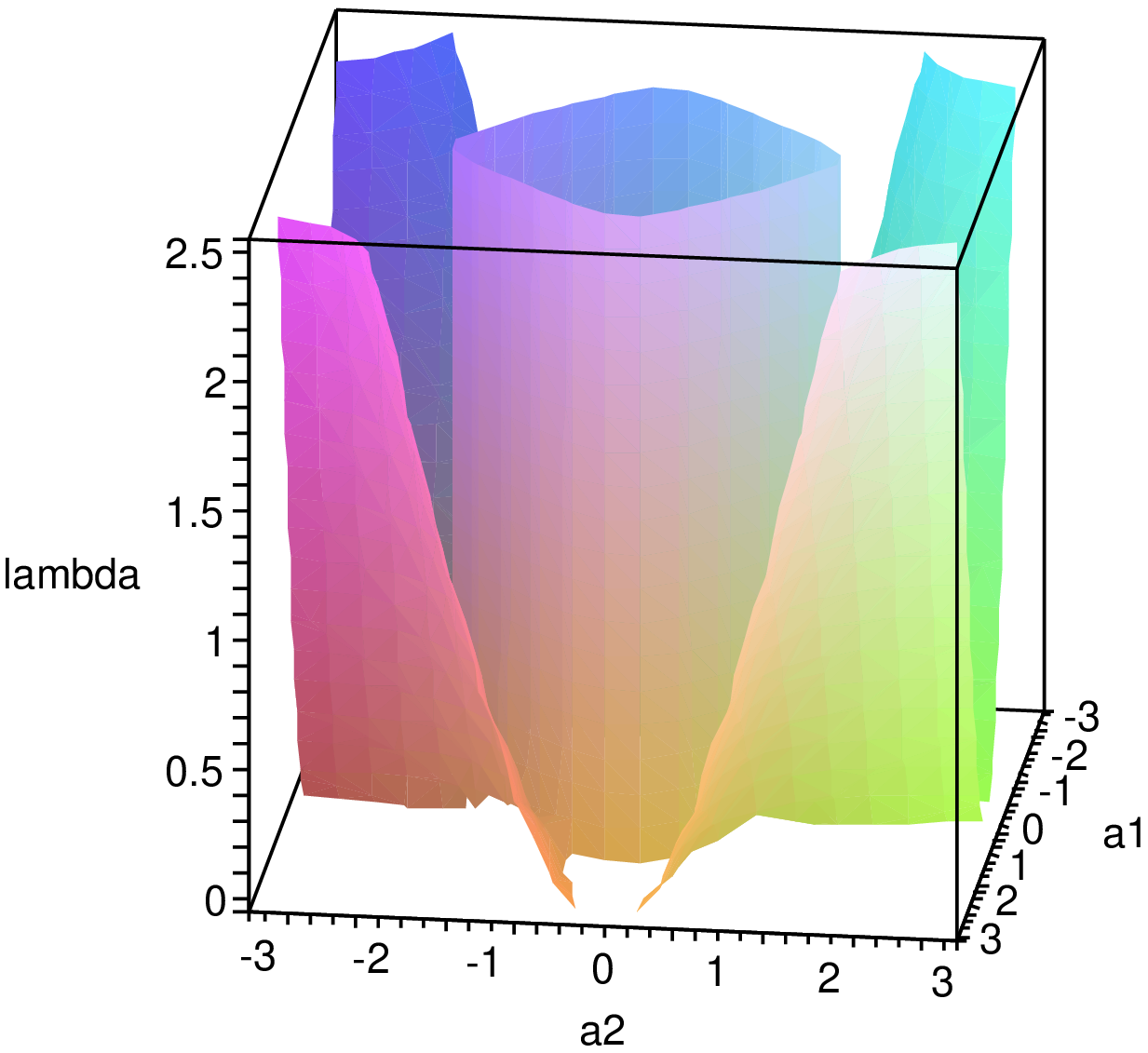}\ 
\includegraphics[width=8cm]{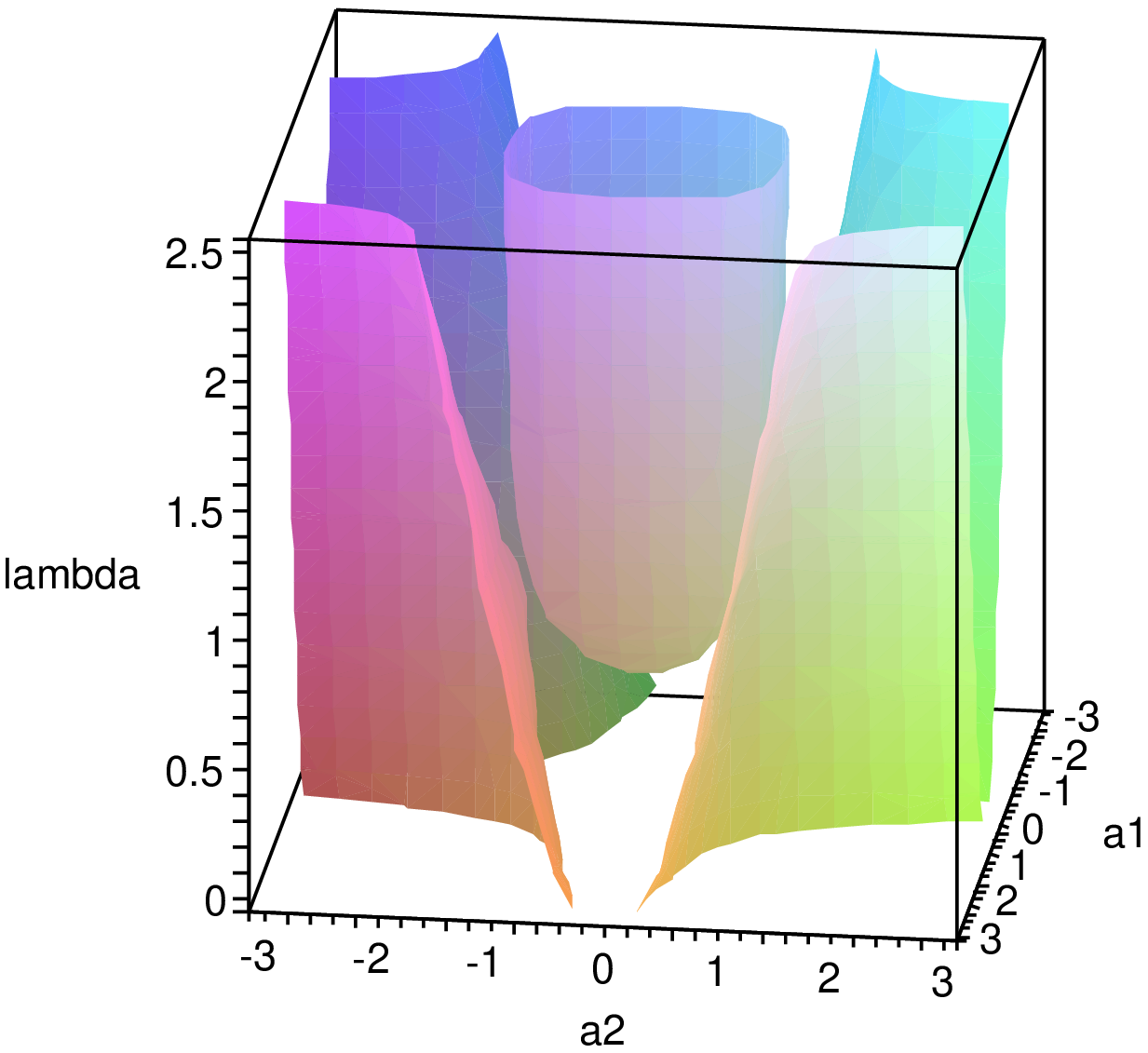}
}
\caption{Surfaces on which $C_\Omega$ changes sign in 5-dimensions.
The left-hand figure shows the locus of zeros and the right hand figure the 
locus of poles.
The left-hand cylinder lies everywhere outside the central pillar in the right-hand figure and everywhere inside the four wings of the right-hand figure,
touching all the right-hand figure surfaces tangentially along the vertical lines $\J_1^2=\J_2^2=1$.
$C_\Omega$  is negative at $\J_1=\J_2=\cosmo=0$ and changes sign every time one of
these surfaces is crossed.}\label{fig:D=5}
\end{figure}

The relevant surfaces for $D=5$ are shown in figure \ref{fig:D=5}.

\begin{figure}[!ht]
\centerline{\includegraphics[width=15cm]{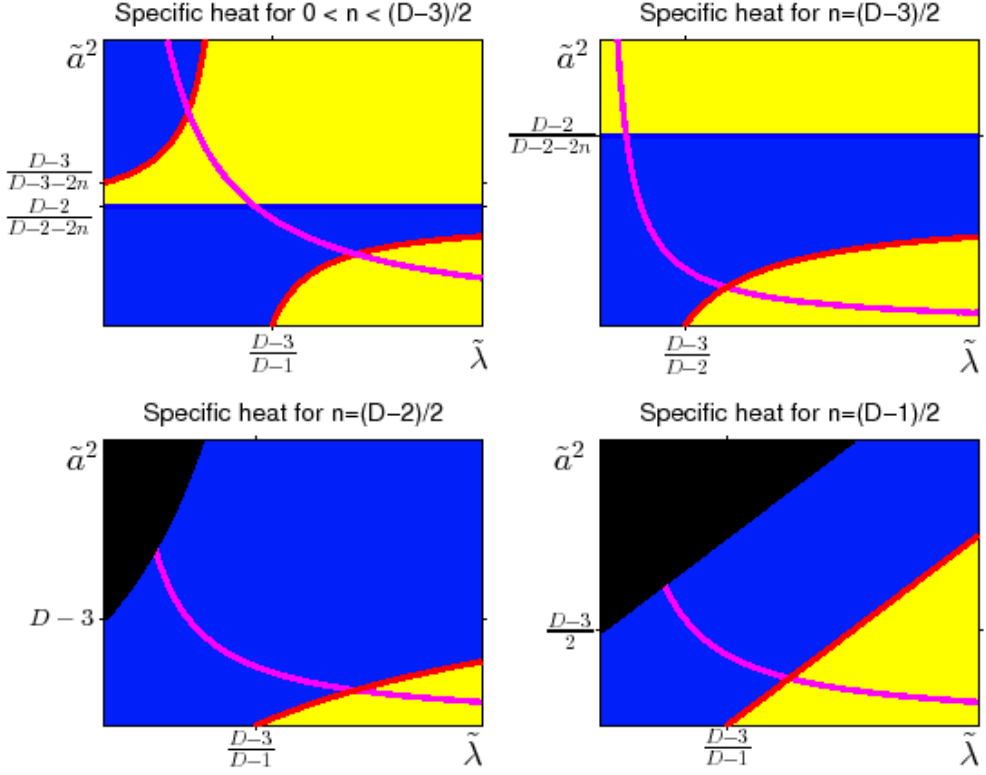}
}
\caption{Regions in the $\J^2-\cosmo$ plane 
where $C_\Omega>0$ are coloured yellow, regions where it is negative are blue
(the generic case is the top-left figure, the other three panels are for special values of $n$, the black regions are where $T<0$). 
$C_\Omega$ is zero on the horizontal lines in the upper two figures and diverges on the red curves.
The geometrical constraint (\ref{eq:Xi-restriction}) 
requires that only the area below the hyperbola $\cosmo\,\J^2=1$ (magenta) is allowed.
}
\label{fig:COmega_equal_J}

\end{figure}

For the symmetric configurations (\ref{eq:n-equal-a})
\beq
\overline \Sigma_1 = \frac{n\J^2}{1-\J^2}
\eeq
and, in the region where $T>0$, (\ref{eq:C_Omega-positivity}) requires
\beq
\frac{D-2 -(D-2-2n)\J^2}{D-3-(D-1)\cosmo -[(D-3-2n)-(D-1-2n)\cosmo]\J^2}<0.
\eeq
$C_\Omega$ can change sign either by passing through zero when the numerator vanishes,
on the $\cosmo$-independent hypersurface
\beq
\J^2=\frac{D-2}{D-2-2n},
\label{eq:zero_surface}
\eeq
or by passing through a pole when the denominator vanishes, 
on the hypersurface
\beq
\J^2=\frac{D-3-(D-1)\cosmo}{D-3-2n-(D-1-2n)\cosmo}.
\label{eq:zeroDen}\eeq
There are no singularities in the range
\beq
\frac{D-3-2n}{D-1-2n} < \cosmo < \frac{D-3}{D-1}
\eeq
The regions in which $C_\Omega$ is positive are shown in figure \ref{fig:COmega_equal_J}. For singly spinning black holes, $n=1$, thermodynamic stability was examined in \cite{DFMS-AdS} and the upper left figure is essentially the same as figure
2 in that reference, expressed in slightly different variables. The analysis
here extends this to multiply spinning black holes.

For $\J^i=0$ (or equivalently $n=0$) a negative cosmological constant of sufficient magnitude, $\cosmo>\frac{D-3}{D-1}$, can render $C_\Omega$ positive, but when $\J^2$ is increased the value of $\cosmo$ necessary to ensure $C_\Omega>0$ must also increase at first, until $\J^2=\frac{D-2}{D-2-2n}$ (for $n\le\frac{D-3}{2}$) at which point $C_\Omega$ becomes positive even for $\cosmo=0$.

The symmetric configurations (\ref{eq:n-equal-a}) do not reveal all of the structure, however.  In $D=5$, for example, 
in addition to the singular branch shown in the bottom right panel of
figure \ref{fig:COmega_equal_J} the denominator of $C_\Omega$ also vanishes 
for $\J_1^2=\J_2^2=1$ for  any $\cosmo$, because there is a factor
$(1-\J^2)$ in the polynomial (\ref{eq:singular_surface}).
This is cancelled by a similar factor 
in the numerator of $C_\Omega$ (\ref{eq:zero_surface})
and $C_\Omega$ is in fact finite on  $\J_1^2=\J_2^2=1$, for any $\cosmo\ne1$.
Away from the symmetric plane $\J_1=\J_2$ however the singular surface bifurcates
and, if $\cosmo>\frac{1}{2}$ (or, more generally, $\cosmo > \frac{D-3}{D-1}$), there are more regions where $C_\Omega$ changes sign. 
In $D=5$, if $\cosmo>\frac{1}{2}$ is fixed and we 
come out from $\J_1=\J_2=0$ in a general
direction in the $\J_1-\J_2$ plane, we start with $C_\Omega>0$, pass through a singularity, then a zero and then a second singularity, so $C_\Omega$ changes sign three times and eventually emerges
with a negative sign.  In the direction $\J_1=\J_2=\J$ however 
one of the singular surfaces and the zero surface  meet so $C_\Omega$ only changes sign once, going straight from positive to negative. The full phase diagram is therefore more involved than shown in figure \ref{fig:COmega_equal_J}.  However when the restrictions arising from positivity of the moment of inertia tensor considered in the next 
subsection are folded into the analysis, the symmetric configurations
(\ref{eq:n-equal-a}) still prove to be sufficient for a good understanding 
of the topology of the local thermodynamically stable region of phase space.

\subsection{Isentropic moment of Inertia}

The isentropic moment of inertia tensor is defined as
$\Inertia^{ij}=\left( \frac{\partial J^i}{\partial \Omega_j}\right)_{S,P}$,
which is derived in appendix \ref{app:I_S}.  It has the form 
\beq
\Inertia^{ij}= 
\frac{r_h S}{2\pi \Xi_i \Xi_j}
\frac{(1+\J_i^2)(1+\J_j^2)}{(1-\J_i^2)(1-\J_j^2)}
\left\{\overline \Xi_i (1-\J_i^2)\delta^{ij} 
- \frac{2(1+\cosmo)\J_i \J_j}{(D-2+2\overline\Sigma_1)}
\right\},\label{eq:IS}
\eeq
where
\beq 
\overline \Xi_i := 1+\lambda a_i^2.
\eeq

The determinant of $\Inertia$ is
\beq \det\Inertia 
=\left(\frac{r_h S}{2\pi} \right)^N 
\frac{\Bigl(D-2-2\cosmo\sum_k\frac{\J_k^2}{\overline \Xi_k}\Bigr)}
{\bigl(D-2+2\overline\Sigma_1\bigr)}
\prod_{i=1}^\dCSA 
\frac{\bigl(1+\J_i^2\bigr)^2 \overline\Xi_i}{\bigl( 1-\J_i^2\bigr) \Xi_i^2}.
\eeq
Hence at least one eigenvalue has a pole on the hypersurface
(\ref{eq:zero_surface}), where $C_\Omega$ has a zero, and there are also
possible poles whenever any $\J_i^2=1$ or $\cosmo\J_i^2=1$, though these last
are never really achieved due to the constraint (\ref{eq:Xi-restriction}). 
The same constraint shows that the determinant never vanishes, since it
implies that $\sum_k\frac{\cosmo\J_k^2}{\overline \Xi_k}<\frac{N}{2}$ so
\beq
D-2-2\cosmo\sum_k\frac{\J_k^2}{\overline \Xi_k}>D-2-N>0.
\eeq
The location of the poles of $\det \Inertia$ are determined purely
by $\J_i$ and are independent of $\cosmo$, and these have 
already been studied
for $\cosmo=0$ in \cite{BPDThermo}.

The eigenvalues do however depend on $\cosmo$, even thought the locus of poles
of $\det\Inertia$ does not.  In the symmetric configurations of the from (\ref{eq:n-equal-a})
the moment of inertia tensor (\ref{eq:IS}) has three distinct eigenvalues,
stripping off the positive pre-factor $\frac{r_hS}{2\pi}$ these are
\bea
\lambda_1&=&\frac{(1+\J^2)^2\bigl(D-2+(D-2-2n)\cosmo\J^2\bigr)}
{\Xi^2\bigl(D-2-(D-2-2n)\J^2\bigr)}\nonumber\\
\lambda_2&=&\frac{(1+\cosmo\J^2)(1+\J^2)^2}{\Xi^2(1-\J^2)},\\
\lambda_3&=&1,\nonumber
\eea
with $\lambda_2$ being $(n-1)$-times degenerate (and hence only present 
for $n\ge 2$) and $\lambda_3$ being $(N-n)$-times degenerate (and only present for $n<N$). 
Both $\lambda_1$ and $\lambda_2$ are positive for small $\J^2$,
but $\lambda_2$ has a pole and changes sign at $\J^2=1$  while
$\lambda_1$ has a pole and changes sign at $\J^2=\frac{D-2}{D-2-2n}$.

Hence the moment of inertia tensor is positive if and only if one of the following conditions holds:
\begin{enumerate}
\item $n=0$;
\item $n=1$ and $0\le \J^2 < \frac{D-2}{D-4}$ (this excludes the region above the horizontal lines in the upper two plots in figure \ref{fig:COmega_equal_J});
\item $2\le n \le N$, positivity of $\lambda_2$ requires $0\le \J^2<1$;
\end{enumerate}
For all $1\le n \le N$, positivity of the moment of inertia tensor combined with positivity of $C_\Omega$ effectively constrains parameter space to lie in the yellow region below the line $\J^2=1$ in figure \ref{fig:COmega_equal_J}.

\section{Conclusions \label{sec:conclusions}}

Necessary and sufficient conditions for local thermodynamical stability of
rotating black holes in asymptotically anti-de Sitter $D$-dimensional space time have been explored.  In asymptotically flat space-time no amount of rotation and/or charge can completely stabilise the system,
but a positive pressure, in the form of a negative cosmological constant 
can make black holes locally thermodynamically stable.

General conditions for local thermodynamical stability of an electrically neutral black hole are:
\begin{itemize}
\item the specific heat at constant $\Omega_i$ and $P$ is positive,
\beq
C_{\Omega,P} = T\left.\frac{\partial S}{\partial T}\right|_{\Omega,P}>0;
\eeq
\item the isentropic moment of inertia tensor
\beq
\Inertia^{ij}
=\left.\frac{\partial J^i}{\partial \Omega_j}\right|_{S,P}
=\left.\frac{\partial J^j}{\partial \Omega_i}\right|_{S,P}
\eeq
is a positive $N\times N$ matrix, where $N$ is the rank of $SO(D-2)$;
\item the adiabatic compressibility is positive,
\beq
\kappa=-\left.\frac{1}{V}\frac{\partial V}{\partial P}\right|_{J^i,S}>0.
\eeq
\end{itemize}
$P=-\frac{\Lambda}{8\pi}$ here is the pressure associated with the cosmological constant and the thermodynamic volume is defined by
\[ V=\left.\frac{\partial H}{\partial P}\right|_{J,S},\]
where the enthalpy $H(J,S,P)$ is equated with the black hole ADM mass, $M$.

For electrically neutral black holes there is a simple relation between the
thermal energy, $U$, the enthalpy, $M$, and the extended free energy,
$\E$, given by equation  (\ref{eq:UEM}).

These conditions have been analysed in detail for asymptotically AdS Myers-Perry
black holes.  Given the complexity of the metric the relevant thermodynamic quantities have remarkably simple expressions when expressed in terms of
the dimensionless variables
$\J^i=\frac{a^i}{r_h}$ and $\cosmo=\frac{2|\Lambda|r_h^2}{(D-1)(D-2)}$.
The extended canonical potential (the Euclidean action times the temperature) 
has the very simple expression (\ref{eq:GrandCan}) in terms of the entropy.

\begin{figure}[!ht]
\centerline{
\includegraphics[width=15cm]{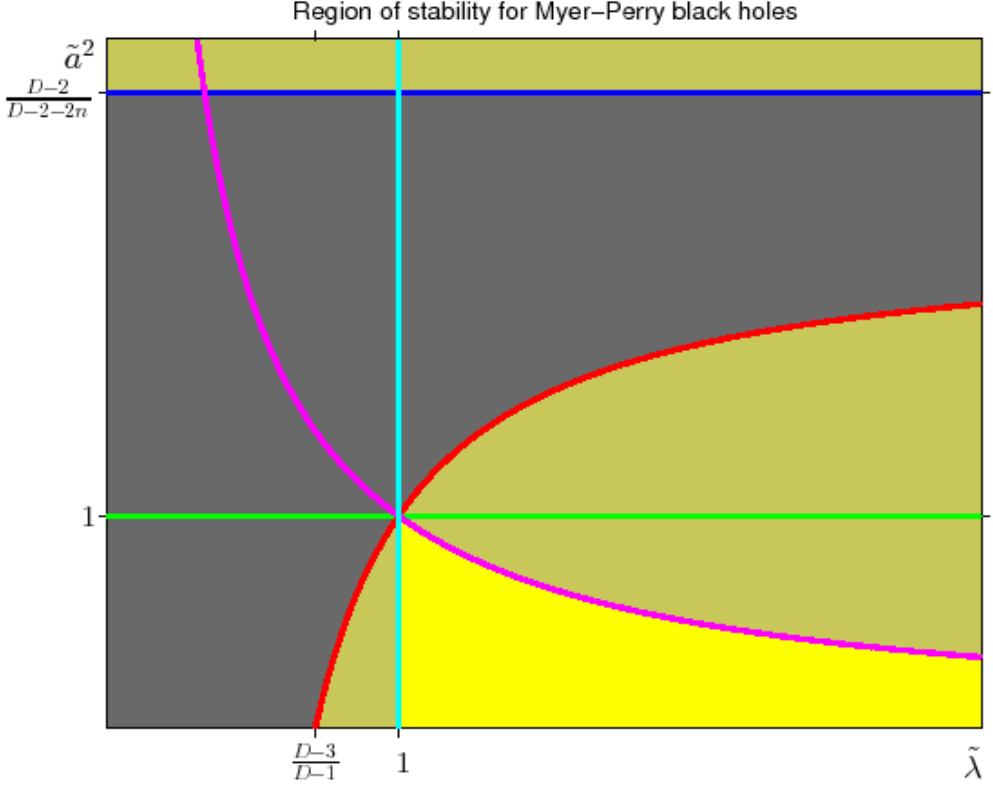}
}
\caption{Region of thermodynamic stability for asymptotically AdS Myers-Perry black holes, for symmetric configurations (\ref{eq:n-equal-a})
with $1<n\le\frac{D-3}{2}$. The thermodynamically stable region is the bright yellow region at the bottom right-hand part of the figure of the figure: to the left of the vertical line at $\cosmo=1$ the black hole is unstable against the Hawking-Page phase transition and the region above the hyperbola (magenta) is excluded by the
geometric constraint (\ref{eq:Xi-restriction}).  The isentropic moment of inertia tensor develops a negative eigenvalue above the horizontal line $\tilde a{}^2=1$ (green) and $C_\Omega$ diverges on the red curve. (For $n=1$ the horizontal green line is moved up to the horizontal blue line at $\tilde a{}^2=\frac{D-2}{D-4}$.)  
For multi-spinning black holes with $n=N$ the region of stability is the same, though some of the other curves change shape.}
\label{fig:Thermo_Stable}

\end{figure}

For symmetric configurations of the form (\ref{eq:n-equal-a})
the region of parameter space $(\J_i,\cosmo)$ for which such black holes
are locally thermodynamically
stable is the bright yellow region in the bottom right of figure \ref{fig:Thermo_Stable}.  
For asymptotically AdS Myers-Perry space-times it is already known that
the adiabatic compressibility is always positive, \cite{BPDCompress-D}.
The singularity structure of the isentropic moment of inertia is independent of $\cosmo$, once the constraint
(\ref{eq:Xi-restriction}) had been taken in to account, and positivity of the eigenvalues merely demands that $\J_i^2<1$, 
except for singly spinning black holes, $n=1$, in which case this is relaxed to $\J^2<\frac{D-2}{D-4}$ (in $D=4$ the isentropic moment of inertia is always positive, though the isothermal moment of inertia is not \cite{Niuetal}).
The remaining requirements for local thermodynamic stability
come from demanding positivity of the specific heat,
$C_\Omega$.

When the condition for stability against Hawking-Page phase transition are folded in, the region of stability is restricted further to the bright yellow region in figure \ref{fig:Thermo_Stable}, lying to the right of $\cosmo=1$.

These thermodynamic constraints must of course be supplemented by possible further
constraints required to avoid the
super-radiant instability of the classical solutions themselves, discussed
in \cite{DFMS-AdS} and \cite{CD}-\cite{SY}. The super-radiant curve for $n=1$ is
$\Omega^2\lambda=1$, which translates in our variables to 
\beq
\J^2=\frac{1}{\cosmo} \qquad \hbox{or} \qquad \J^2= \cosmo,
\label{eq:SuperRad}\eeq
and the local thermodynamic stability conditions $\J^2<1/\tilde \lambda$ with $\tilde\lambda>1$
ensure that the black hole will not super-radiate.
 
It would be of interest to extend this analysis to include electrically charged black holes, although from the discussion in section \S\ref{sec:GeneralStability} this 
cannot change the fact that such black holes are locally thermodynamically unstable
for $\Lambda=0$  
when unconstrained, the $\Lambda<0$ case would introduce another dimension into
the phase diagram.  One should also not forget that the thermodynamic stability of various constrained ensembles would be of interest, while there has already been a substantial amount of work on this  for $Q\ne 0$ and $J^i=0$ a complete analysis including both in any $D$ is not yet available.  The considerations here could also be applied
to other rotating solutions in higher dimensions with event horizon topologies that differ from spheres, such as black rings and black saturns \cite{BR1}-\cite{BR7}, although
there are no exact solutions of this form currently known 
in the asymptotically AdS case there are approximate 
solutions in certain regimes of parameter space.
 
\begin{appendix}

\section{General stability criteria \label{app:Stability}}

In this appendix we examine the criteria for thermodynamic stability for
electrically neutral black holes in some
detail.
We first show that the Hessian (\ref{eq:WAB}) can be partially diagonalised by changing variables from $X^A=(J^i,S,V)$, with $A=1,\ldots,N+2$,
to \beq X^{A'}=(X^1,\ldots,X^{\dCSA+1},x_{\dCSA+2}),\eeq
so, in particular,  $X^{\dCSA +2'}=x_{\dCSA+2}=\left.\frac{\partial U}{\partial V}
\right|_{X^a}=-P$ 
and $X^{a'}=X^a$ for
$a=1,\ldots,\dCSA+1$.
The stability properties can equally well be explored in these co-ordinates: 
although the magnitude of the eigenvalues of the Hessian may change under this co-ordinate 
transformation, the signature does not, so $\U_{AB}=\frac{\partial^2 U}{\partial X^A \partial X^B}$ is a positive matrix
if and only if $\U_{A'B'}$ is, provided the co-ordinates transformation is not
singular.

Let $a,b=1,\ldots \dCSA+1$ and 
\beq \xi_a=\frac{\partial^2 U
}{\partial X^a \partial X^{\dCSA+2}}
=\left.\frac{\partial x_{\dCSA+2}}{\partial X^a}\right|_{X^{\dCSA+2}}
=\left.\frac{\partial x_a}{\partial X^{\dCSA+2}}\right|_{\vec{X}},\eeq
where $\vec{X}=(X^1,\ldots,X^{\dCSA+1})$.  Then
\beq 
\U_{AB}= \begin{pmatrix}
\U_{ab}  & \xi_a \\ 
\xi_b  & \U_{\dCSA+2,\dCSA+2} 
\end{pmatrix}.
\eeq
Changing variables from $X^A$ to $X^{A'}$,
\beq
\frac{\partial X^{A'}}{\partial X^{B}}
=\begin{pmatrix}
\delta_{ij} & {\bf 0} & {\bf 0} \\ {\bf 0}^t & 1 & 0 \\
\xi_j & \xi_{N+1} & -\left.\frac{\partial P}{\partial V}\right|_{J,S}
\end{pmatrix}
\eeq
with $i,j=1,\ldots,N$  and ${\bf 0}$ an $N$-component column of zeros,
results in
\beq
\U_{AB}dX^A d X^B =  \U_{A'B'}dX^{A'} d X^{B'}
\eeq 
where
\beq
\U_{A' B'}=
\begin{pmatrix}
\U_{ab}  - \bigl(\U_{\dCSA+2,\dCSA+2}\bigr)^{-1}\xi_a \xi_b & 0\\ 
0  & \bigl(\U_{\dCSA+2,\dCSA+2}\bigr)^{-1} 
\end{pmatrix}.\label{eq:Wchi}
\eeq
This expression can be further simplified by noting that 
\bea
\U_{ab}=\left.\frac{\partial x_b}{\partial X^a}\right|_{X^{\dCSA+2}}&=&
\left.\frac{\partial x_b}{\partial X^a}\right|_{x_{\dCSA+2}}+
\left.\frac{\partial x_b}{\partial x_{\dCSA+2}}\right|_{\vec{X}}
\left.\frac{\partial x_{\dCSA+2}}{\partial X^a}\right|_{X^{\dCSA+2}}\nonumber\\
&=&
\left.\frac{\partial x_b}{\partial X^a}\right|_{x_{\dCSA+2}}+
\left.\frac{\partial x_b}{\partial X^{\dCSA+2}}\right|_{\vec{X}}
\left.\frac{\partial X^{\dCSA+2}}{\partial x_{\dCSA+2}}\right|_{\vec{X}}
\left.\frac{\partial x_{\dCSA+2}}{\partial X^a}\right|_{X^{\dCSA+2}}\nonumber\\
&=&
\left.\frac{\partial x_b}{\partial X^a}\right|_{x_{\dCSA+2}}+
\bigl(\U_{\dCSA+2,\dCSA+2}\bigr)^{-1}\xi_a\xi_b\, ,\label{eq:WxX}
\eea
thus
\beq
\U_{ab}-\bigl(\U_{\dCSA+2,\dCSA+2}\bigr)^{-1}\xi_a \xi_b=
\left.\frac{\partial x_b}{\partial X^a}\right|_{x^{\dCSA+2}}.
\eeq
Denote this $(\dCSA +1)\times (\dCSA +1)$ matrix by $\widetilde \U$,
with components
\beq 
\widetilde{\U}_{ab} = \left.\frac{\partial x_b}{\partial X^a}\right|_{x^{\dCSA+2}}=\left.\frac{\partial^2 H}{\partial X^a \partial X^b}\right|_{x_{N+2}}
\eeq
where $H=U+PV=M$ is the enthalpy.
Then $\U$ is partially diagonalised in the $X^{A'}$ co-ordinates
\beq
\U_{A'B'}=
\begin{pmatrix}
 \widetilde{\U}_{ab} & 0 \\
0 & \bigl(\U_{\dCSA+2,\dCSA+2} \bigr)^{-1}
\end{pmatrix}.
\eeq
Now
\beq
\U_{\dCSA+2,\dCSA+2}=-\left.\frac{\partial P}{\partial V}\right|_{J^i,S}
\eeq
which is related to the adiabatic compressibility
\beq
\kappa_{J,S}=-\left.\frac{1}{V}\frac{\partial V}{\partial P}\right|_{J^i,S}
\eeq
by
\beq
\U_{\dCSA+2,\dCSA+2}= \frac{1}{\kappa_{J,S} V}
\eeq
so
\beq
\U_{A'B'}=
\begin{pmatrix}
 \widetilde{\U}_{ab} & 0 \\
0 & \kappa_{J,S} V
\end{pmatrix}.
\eeq
Assuming that adiabatic compressibility and the volume are positive
(which is known to be the case for asymptotically AdS Myers-Perry black holes
in $D$-dimensions, \cite{BPDCompress-D}), 
the question of thermodynamic stability has now been reduced to the question
of positivity of $\widetilde \U$.  In terms of familiar thermodynamic quantities
$\widetilde \U$ decomposes as
\beq \widetilde\U= \begin{pmatrix}
 (\Inertia_S^{-1})_{ij} & \zeta_i  \\
\zeta_j & \frac{1}{\beta C_{J,P}}  
\end{pmatrix},
\eeq
where $\Inertia_S^{ij}=\left.\frac{\partial J^i}{ \partial \Omega_j}\right|_{S,P}$
is the isentropic moment of inertia tensor, $\beta = \frac{1}{T}$,
$C_{{\bf J},P}$ is the heat capacity at constant angular 
momentum and pressure and 
\beq\zeta_i=
\left.\frac{\partial T}{\partial J^i}\right|_{S,P}
=\left.\frac{\partial \Omega_i}{\partial S}\right|_{J,P}
\label{eq:zetaMaxwell}\eeq
(equation (\ref{eq:zetaMaxwell}) is a Maxwell relation).

We can try to continue the process of partial diagonalisation and make a 
further coordinate transformation to
\beq X^{A''}=(X^1,X^2,\ldots,X^N,x_{N+1},x_{N+2}) \eeq
with $x_{N+1}=T$, $x_{N+2}=-P$ and $X^{i''}=X^{i'}=X^i$ for $i=1,\ldots,N$.
\beq
\frac{\partial X^{A''}}{\partial X^{B'}}=
\begin{pmatrix}
\delta_{ij} & {\bf 0} & {\bf 0} \\ 
\sigma_j & \sigma_{N+1} & \sigma_{N+2}\\
{\bf 0}^t & 0 & 1 \\
\end{pmatrix}
\eeq
where
\beq \sigma_i = \left.\frac{\partial T}{\partial J^i}\right|_{S,P}
=\left.\frac{\partial \Omega_i}{\partial S}\right|_{J,P},
\quad \sigma_{N+1}=\left.\frac{\partial T}{\partial S}\right|_{J,P} 
\quad\hbox{and}\quad
\sigma_{N+2}=-\left.\frac{\partial T}{\partial P}\right|_{J,S}.
\eeq
In these variables
\beq
\U_{A''B''}=\begin{pmatrix}
\widetilde{\widetilde\U}_{ij} & {\bf 0} & {\bf 0} \\
{\bf 0}^t & \beta C_{J,P} & -\beta C_{J,P}\,\sigma_{N+2}  \\
{\bf 0}^t &  -\beta C_{J,P}\,\sigma_{N+2}  &  \kappa_{J,S} V+ \beta C_{J,P}\, \sigma^2_{N+2}
\end{pmatrix}
\label{eq:WJTP}
\eeq
with
\beq
\widetilde{\widetilde\U}_{ij}=\widetilde\U_{ij} - \beta C_{J,P}\,\sigma_i\sigma_j.\label{eq:WWss}
\eeq
In fact $\widetilde{\widetilde\U}$ is the inverse of the isothermal moment of
inertia tensor, $\Inertia_T$.
To see this note that
\beq \bigl(\Inertia_T\bigr)^{ij} 
= \left.\frac{\partial J^i}{\partial \Omega_j}\right|_{T,P},
\eeq
and $\Omega_i(X^{A''})=\Omega_i(J,T,P)$ so
\bea \widetilde\U_{ij}&=&
\left.\frac{\partial \Omega_i}{\partial J^j}\right|_{S,P}
=\left.\frac{\partial \Omega_i}{\partial J^j}\right|_{T,P}+
\left.\frac{\partial \Omega_i}{\partial T}\right|_{J,P}
\left.\frac{\partial T}{\partial J^i}\right|_{S,P}\nonumber\\
&=&\left.\frac{\partial \Omega_i}{\partial J^j}\right|_{T,P}+
\left.\frac{\partial \Omega_i}{\partial S}\right|_{J,P}
\left.\frac{\partial S}{\partial T}\right|_{J,P}
\left.\frac{\partial T}{\partial J^i}\right|_{S,P}\nonumber\\
&=&(\Inertia_T^{-1})_{ij} + \beta C_{J,P}\sigma_i\sigma_j,
\eea
since 
\beq C_{J,P}= T\left.\frac{\partial S}{\partial T}\right|_{J,P},\eeq
and the result follows from (\ref{eq:WWss}).

So we can write (\ref{eq:WJTP}) as
\beq
\U_{A''B''}=\begin{pmatrix}
(\Inertia_T^{-1})_{ij} & {\bf 0} & {\bf 0} \\
{\bf 0}^t & \beta C_{J,P} & -\beta C_{J,P}\,\sigma_{N+2}  \\
{\bf 0}^t &  -\beta C_{J,P}\,\sigma_{N+2}  &  \kappa_{J,S} V+ \beta C_{J,P}\, \sigma^2_{N+2}
\end{pmatrix}.
\eeq

The physical conditions for complete local thermodynamic stability are now
clear: the isothermal moment of inertia tensor $\Inertia_T$
must be a positive matrix, which of course ensures that its inverse is also
positive; and in addition the $2\times 2$ matrix
\beq\begin{pmatrix}
\beta C_{J,P} & -\beta C_{J,P}\,\sigma_{N+2}  \\
 -\beta C_{J,P}\,\sigma_{N+2}  &  \kappa_{J,S} V+ \beta C_{J,P}\, \sigma^2_{N+2}
\end{pmatrix}\eeq 
must be positive.  The determinant of this matrix is just
$\beta \kappa_{J,S} C_{J,P} V$, so positivity is ensured by demanding that
the isentropic compressibility $\kappa_{J,S}$ is positive and that the heat capacity $C_{J,P}$ is positive. 

The final conclusion is that local thermodynamic stability holds 
if and only if the following three perfectly reasonable conditions hold
(assuming that $\beta$ and $V$ are positive):
\begin{itemize}
\item $\kappa_{J,S}>0$;
\item  $C_{J,P}>0$;
\item $\Inertia_T$ is a positive matrix.
\end{itemize}
At fixed pressure positivity of $C_J$ and $\Inertia_T$ is equivalent to
positivity of the specific heat at constant angular velocity, $C_\Omega$, and positivity of the isentropic moment of inertia tensor, $\Inertia_S$. 
This was proven 
in \cite{BPDThermo} 
for $P=0$ and the same analysis goes through for any $P\ge 0$. 
The essential idea of the proof is to Legendre transform $H(J^1,\ldots,J^N,S,P)$
to 
\beq G(\Omega_1,\ldots,\Omega_N,T,P)=H-TS - {\mathbf \Omega}.{\mathbf J}\eeq
Now $(\Inertia_S^{-1})_{ij}=\frac{\partial^2 H}{\partial J^i \partial J^j}$ and $\frac{1}{\beta C_{J,P}}=\frac{\partial^2 H}{\partial S^2}$ are components of the Hessian of $H$
while $(\Inertia_T)_{ij}=-\frac{\partial^2 G}{\partial \Omega_i \partial \Omega_j}$ and $\beta C_{\Omega,P}=-\frac{\partial^2 G}{\partial T^2}$ are components of the Hessian of $-G$. As matrices these Hessians are inverses of each other, so
positivity of one ensures positivity of the other. 
Indeed the identity
\beq
\beta C_J \det(\Inertia_T) = \beta C_\Omega \det(\Inertia_S), 
\eeq
proven in \cite{BPDThermo} for $P=0$, also holds for $P>0$.
 
An equivalent set of conditions for complete local thermodynamic stability 
is therefore that the following conditions hold
(again assuming that $\beta$ and $V$ are positive):
\begin{itemize}
\item $\kappa_{J,S}>0$;
\item  $C_{\Omega,P}>0$;
\item $\Inertia_S$ is a positive matrix.
\end{itemize}
In the text the example of asymptotically anti-de Sitter Myers-Perry black holes
is treated in detail and it transpires that it is much easier to calculate
$\Inertia_S$ and $C_{\Omega,P}$ for these metrics in $D$-dimensional space-time
than it is to calculate $\Inertia_T$ and $C_{J,P}$. In the text we therefore
focus on the former, in particular only $\Inertia_S$ is considered and the
subscript $S$ is dropped, $\Inertia = \Inertia_S$.

\section{Specific heat at constant angular velocity\label{app:COmega}}

The specific heat at constant $\Omega$ is straightforward to determine, using the
same kind if calculations as those in \cite{BPDThermo} 
for the asymptotically flat case.  
In terms of the dimensionless quantities
\beq \J_i := \frac {a_i}{r_h} \eeq
and 
\beq\Xi_i := 1- \lambda a_i^2 =1-\cosmo \J_i^2 \label{eq:Xidef} \eeq
we have 
\beq
S=\frac{\varpi r_h^{D-2}}{4}\prod_{i=1}^\dCSA\frac{(1+\J_i^2)}{\Xi_i},
\label{eq:lambdaSdef}\eeq
\beq
T=\frac{1}{4\pi r_h}\left\{
D-3-2 \Sigma_1 + \cosmo(D-1-2\Sigma_1)\right\}\label{eq:lambdaTdef}
\eeq
\beq \Omega_i=\frac{(1+\cosmo) \J_i}{r_h(1+\J_i^2)}.\label{eq:lambdaOmegadef}\eeq
where $\Sigma_1:=\sum_{i=1}^\dCSA\frac{\J_i^2}{1 + \J_i^2}$.

The specific heat at constant angular velocity and pressure is defined as
\beq C_\Omega = T\left( \frac{\partial S}{\partial T}\right)_{\Omega,\lambda}. \eeq
From (\ref{eq:lambdaOmegadef}) we find
\beq \Omega_i=const \qquad\Rightarrow \qquad 
\left.d\J_i\right|_{\Omega,\lambda} = \left.\left(\frac{1-\cosmo}{1+\cosmo}\right)
\left(\frac{1+\J_i^2}{1-\J_i^2}\right)\J_i \frac{d r_h}{r_h}\right|_{\Omega,\lambda} \eeq
and from (\ref{eq:Xidef})
\beq \left.\frac{d\Xi_i}{\Xi_i}\right|_{\Omega,\lambda} = - \frac{4\cosmo \J_i^2}{(1+\cosmo)(1-\J_i^2)}\left.\frac{d r_h}{r_h}\right|_{\Omega,\lambda}. \eeq

Using these it is straightforward to show that
\beq
\left. \frac{\partial T}{\partial r_h}\right|_{\Omega,\lambda} = -\frac{1}{4\pi r_h^2}
\left\{D-3 + 2\overline\Sigma_1 -\cosmo (D-1 + 2\overline\Sigma_1)\right\},
\eeq
where $\overline\Sigma_1:=\sum_{i=1}^\dCSA\frac{\J_i^2}{1 - \J_i^2}$,
and
\beq
\left. \frac{\partial S}{\partial r_h}\right|_{\Omega,\lambda} = 
\frac{S}{r_h} (D-2 + 2\overline\Sigma_1).
\eeq
Combining these we immediately arrive at equation (\ref{eq:C_Omega}) in the text,
\beq C_\Omega=  -\frac{4\pi r_h T S (D-2+2\overline\Sigma_1)}
{D-3 + 2\overline\Sigma_1 -\cosmo (D-1 + 2\overline\Sigma_1)}.
\eeq
This generalises the asymptotically anti-de Sitter $D=4$ case derived in 
\cite{MPS-AdS} to arbitrary $D$. It also generalises the asymptotically flat $D$-dimensional case derived in \cite{BPDThermo} to asymptotically AdS.

\section{Isentropic moment of inertia \label{app:I_S}}

To calculate the isentropic moment of inertia we can first use (\ref{eq:lambdaSdef}) to obtain
\bea
d r_h \bigr|_{S,\lambda} &=& -\frac{2 r_h(1+\cosmo)}{(D-2+2\cosmo X)} \sum_{j=1}^\dCSA \left.\frac{\J _j\, d\J_j}{(1+\J_j^2)\Xi_j}\right|_{S,\lambda}\nonumber \\
\Rightarrow \qquad 
\left.\frac{1}{r_h}\frac{\partial r_h}{\partial \J_j} \right|_{S,\lambda} &=& 
-\frac{2 (1+\cosmo)}{(D-2+2\cosmo X)} 
\frac{\J _j}{(1+\J_j^2)\Xi_j}
\eea
with $ X:= \sum_{k=1}^\dCSA \frac{\J_k^2}{\Xi_k}$.
Then
\beq \Omega_i = \frac{\J_i}{r_h(1+\J_i^2)}
\eeq
results in 
\beq
\left.\frac{\partial \Omega_i}{\partial \J_j}\right|_{S,\lambda}
=\frac{(1+\cosmo)}{r_h (1+\J_i^2)(1+\J_j^2)}
\left[(1-\J_i^2) \delta_{ij}+ 2 \frac{(1-\cosmo)\J_i \J_j}{(D-2+2\cosmo X)\Xi_j} \right] .
\eeq
This can be inverted as a matrix to give
\beq
\left.\frac{\partial \J_i}{\partial \Omega_k}\right|_{S,\lambda}
=\frac{r_h}{(1+\cosmo)}\frac{(1+\J_i^2)(1+\J_k^2)}{(1-\J_i^2)(1-\J_k^2)}
\left[(1-\J_i^2) \delta_{ik}
- 2 \frac{(1-\cosmo)\J_i \J_k}{(D-2+2\overline \Sigma_1)\Xi_k} \right].
\label{eq:djdOmega} \eeq

Furthermore 
\beq J_i=(1+\cosmo) \frac{S}{2\pi}\frac{\J_i}{\Xi_i}\eeq
yields
\beq
\left. \frac{\partial J_i}{\partial \J_k}\right|_{S,\lambda}
=(1+\cosmo) \frac{S}{2\pi \Xi_i^2}
\left[\overline \Xi_i \delta_{ik} 
-\frac{4\cosmo}{(D-2+2\cosmo X)}\frac{\J_i\J_k(1+\J_i^2)}
{\Xi_k(1+\J_k^2)}\right].\label{eq:dJda}
\eeq
Equations (\ref{eq:djdOmega}) and (\ref{eq:dJda}) can now be combined to give
the result quoted in the text,
\beq
\Inertia^{ij}= \frac{r_h S}{2\pi}\frac{1}{\Xi_i\Xi_j}
\frac{(1+\J_i^2)(1+\J_j^2)}{(1-\J_i^2)(1-\J_j^2)}
\left\{\overline \Xi_i (1-\J_i^2)\delta_{ij} - \frac{2(1+\cosmo)\J_i \J_j}{(D-2+\overline\Sigma_1)}
\right\}.
\eeq
This extends the asymptotically flat result in \cite{BPDCompress-D} to $\lambda\ne 0$.

The determinant is
\beq
\det(\Inertia)
=\left(\frac{r_h S}{2\pi} \right)^\dCSA
\frac{(D-2-2\cosmo \overline X)}{(D-2+2\overline\Sigma_1)}
 \prod_{k=1}^\dCSA \frac{(1+\J_k^2)^2 \overline\Xi_k}{(1-\J_k)\Xi_k^2},
\eeq
where
\beq
\overline X:= \sum_{k=1}^\dCSA \frac{\J_k^2}{\overline \Xi_k}.
\eeq

Finally we note that there is a cancellation of factors in the combination 
\beq
\beta C_\Omega \det(\Inertia)=  -8\pi^2 \left(\frac{r_h S}{2\pi} \right)^{\dCSA+1}
\kern -10pt \frac{(D-2-2\cosmo \overline X)}
{\bigl[D-3 + 2\overline\Sigma_1 -\cosmo (D-1 + 2\overline\Sigma_1)\bigr]}
 \prod_{k=1}^\dCSA \frac{(1+\J_k^2)^2 \overline\Xi_k}{(1-\J_k)\Xi_k^2}.
\eeq
\end{appendix}

\end{document}